\documentclass[aps,prb,preprint,preprintnumbers,amsmath,amssymb]{revtex4-2}
\usepackage{graphicx}
\usepackage{epsfig}
\usepackage{color}
\usepackage{dcolumn}
\usepackage{bm}
\usepackage{float}
\usepackage{amsmath}
\usepackage{gensymb}
\DeclareMathAlphabet{\mathbbmsl}{U}{bbm}{m}{sl}
\makeatletter
\newsavebox{\@brx}
\newcommand{\llangle}[1][]{\savebox{\@brx}{\(\m@th{#1\langle}\)}%
	\mathopen{\copy\@brx\kern-0.5\wd\@brx\usebox{\@brx}}}
\newcommand{\rrangle}[1][]{\savebox{\@brx}{\(\m@th{#1\rangle}\)}%
	\mathclose{\copy\@brx\kern-0.5\wd\@brx\usebox{\@brx}}}
\makeatother
\begin{document}
\draft

\title{Flat bands of TaS$_2$ under superlattice potential modulation: A Wannier tight-binding model study}

\author{Thi-Nga Do$^{1}$ and Godfrey Gumbs$^{2}$
}
\affiliation{$^{1}$ Department of Physics and Astronomy, The University of Tampa, Tampa, Florida 33606, USA  \\
$^{2}$ Department of Physics and Astronomy, Hunter College of the City University of New York,
695 Park Avenue, New York, New York 10065, USA
}

\date{\today}

\begin{abstract}
In this work, we construct a Wannier tight-binding model for TaS$_2$ under a superlattice potential modulation, based on the Joint Automated Repository for Various Integrated Simulations database established by the U.S. National Institute of Standards and Technology, so as to study the electronic properties of the structure. Our computational method enables direct calculation of the energy bands from the Hamiltonian without any additional assumptions. We observed a pair of dispersionless flat bands, significant interactions between energy bands, and nontrivial modification of band dispersion at low modulated electric potentials. This work provides a valuable reference for researchers investigating two-dimensional condensed matter materials under superlattice potential modulation.
\end{abstract}
\pacs{PACS:}
\maketitle

\section{Introduction}

Flat energy band structure is an important topic in condensed matter physics due in part to its intriguing properties. These phenomena include the quantized anomalous Hall effect \,\cite{QAHE, QAHE1}, superconductivity \cite{superconductor, superconductor1}, and charge-ordered states \cite{charge, charge1}. The emergence of flat bands can be attributed to several factors such as unique lattice geometries, topological effects, strong orbital hybridization, and electron-electron interactions. Recent studies have demonstrated that flat bands can also be engineered through superlattice potential modulation. For example, flat minibands can form in Bernal-stacked bilayer graphene under a honeycomb-lattice-shaped potential \cite{bigraphene, bigraphene1}. This modulated electric potential can also control phase transitions between quantum anomalous Hall insulator, trivial insulator, and metallic states. With recent experimental advances, artificial superlattices can now be fabricated using techniques like gate-patterning \cite{gate1, gate2} and by exploiting adjacent $moir\acute{e}$  materials \cite{exploi1, exploi2}, with their physical implications being widely explored.

\medskip
\par

Transition metal dichalcogenides (TMDCs) have gained significant attention as promising two-dimensional (2D) materials due to their exceptional electronic, optical, mechanical, and transport properties, making them attractive for various applications in electronics, optoelectronics, and energy storage. Among these, group-V TMDCs, such as TaS$_2$, have shown particular promise for use in transparent conductors, sensors, and atomically thin interconnects \cite{application1, application2}. Various experimental techniques have enabled the successful synthesis of TaS$_2$ monolayers, including chemical vapor deposition \cite{cvp}, molecular beam epitaxy \cite{epitaxy}, and mechanical exfoliation \cite{exfo}. So far, both theoretical and experimental research has revealed a variety of fascinating phenomena in TaS$_2$, covering superconductivity \cite{superconduct}, charge density waves \cite{chargedensity},ferromagnetic states \cite{ferro}, and various phase transitions \cite{phase}.

\medskip
\par

The periodically modulated density of electric charge in 1T-TaS$_2$ by means of charge-density wave at high temperature has been previously discussed \cite{tas2flat, tas2flat1, tas2flat2}. This results in a robust isolated flat band. This has motivated us to explore the electronic structures of TaS$_2$ under a superlattice potential modulation and to search for the possible existence of flat bands. Here, we present a Wannier tight-binding model study of TaS$_2$ in a periodically modulated electric potential and investigate the relation to details of the band structure. Our results show that such a superlattice potential not only generates flat bands at low energies but also induces band interactions and significantly alters energy dispersion.

\medskip
\par

\section{Theoretical method}

The Wannier tight-binding model Hamiltonian (WTBH) has proven to be a powerful tool for investigating physical phenomena of solid-state materials. For example, the electronic properties, optical spectra, magnetic quantization, and the quantum Hall conductivity. The advantage lies in the numerical computation on a relatively coarse real-space grid based on the localization of Wannier functions. Here, we employ the Joint Automated Repository for Various Integrated Simulations (JARVIS) WTBH database established by the U.S. National Institute of Standards and Technology (NIST) \cite{jarvis} for the band structure calculation of TaS$_2$. The JARVIS WTBH database includes thousands of materials, including experimentally synthesized and theoretically predicted ones. The accuracy of this method was evaluated by comparing the Wannier band structures to those calculated directly by high-throughput density functional theory (DFT) method. We have developed a Matlab code to load and parse the JARVIS WTBH and directly export the energy bands. This is a critical step for the next phase of numerical computation where several factors are taken into account.

\medskip
\par

We investigate the electronic properties of 1T-TaS$_2$ monolayer under the effect of a modulated electric potential. This TMDC system is composed of a transition metal Ta and a chalcogen S \cite{tas2model}. The lattice structure of 1T-Ta$_2$ systems is presented in Figs. 1(a) and 1(b) with both top and side views. It presents the sandwich structure consisting of a middle layer, made of Ta atoms, and two sides, made of S atoms. This 1T phase can be considered as a structure composed of an ABC stacking manner and being centrosymmetric. Theoretical internal coordinates are shown in Fig. 1(c). The lattice constant $a$ = 3.38 \AA, $h_S$ = 1.53 \AA, and the angles $\alpha = 94^\circ$ and $\beta = 86^\circ$.

\medskip
\par

\begin{figure}[h]
\centering
{\includegraphics[width=0.9\linewidth]{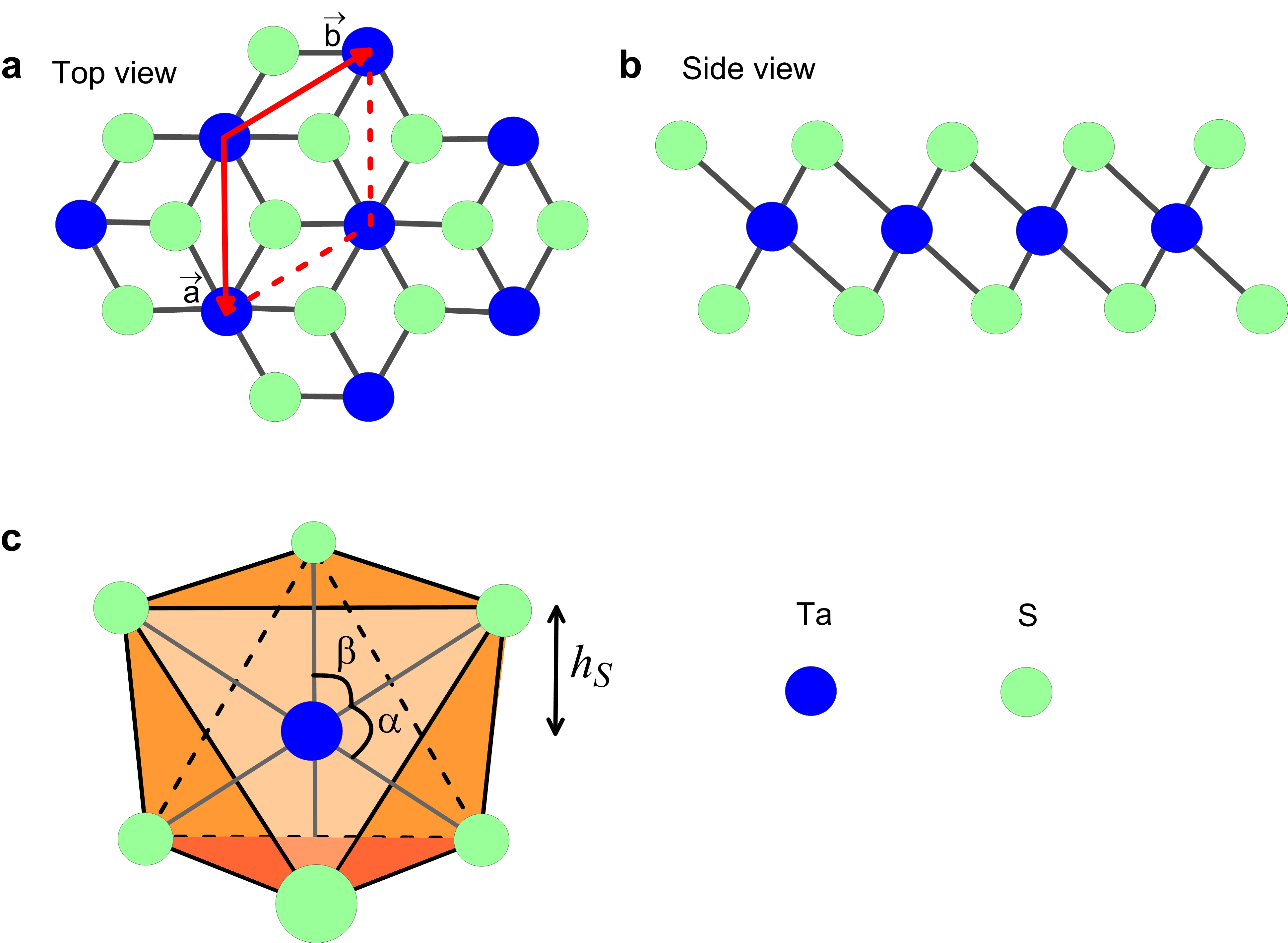}}
\caption{ {\bf a,} Top view and {\bf b,} side view of 1T-TaS$_2$ lattice structure. {\bf c,} First Brillouin zone with high symmetry points.}
\label{Fig1}
\end{figure}

It was shown for pristine graphene that modifying the lattice with a smooth periodic potential can alter the band structure through folding of the Dirac cone into mini bands \cite{folding}. When TaS$_2$ is subjected to a periodically applied electric potential, the crystal's unit cell undergoes a periodic expansion. The extent of this expansion is directly determined by the period of the applied potential. This modulation induces a hexagonal lattice structure, as depicted in Fig. 3, where the periodicity of the potential is described by the factor $n \times n$, with $n$ being a nonzero integer. Essentially, the unit cell is enlarged to match the periodicity of the hexagonal potential. This superlattice potential introduces a modification of the system's Hamiltonian, represented by a term $H_{SL}$, which can be written as \cite{Hsl}

\begin{equation}
H_{SL} = 2W\sum_{i} \cos({\bf G}_i.{\bf r}).
\end{equation}
In this notation, ${\bf G}_i$ are lattice vectors in momentum space which can be expressed as ${\bf G}_1 = \frac{2\pi}{3L}(3,\sqrt{3})$, ${\bf G}_2 = \frac{2\pi}{3L}(0,-2\sqrt{3})$, and ${\bf G}_3 = {\bf G}_1 + {\bf G}_2$. They are parametrized by an effective energy constant $W$. Additionally, $L$ is the lattice constant for the superlattice. The addition of $H_{SL}$ accounts for the effects due to the periodic modulation on the electronic structure, influencing the behavior of the band structure and the formation of new electronic states. This allows for a deeper understanding of the way in which the periodic potential alters the material's properties, leading to phenomena such as flat bands and band inversions.

\medskip
\par

\section{Results and Discussion}

The band structure for TaS$_2$ along high-symmetry directions is shown in Fig. 2(a). Our calculated energy bands are in good agreement with the DFT results presented in Ref. \cite{tas2model}. A clear band gap exists between the valence and conduction bands across the entire first Brillouin zone. Both the valence and conduction bands exhibit multiple extrema at the M, K, and $\Gamma$ points. Notably, the conduction and valence bands approach each other at the $\Gamma$ point, resulting in band degeneracy. We also observe saddle points near the $\Gamma$ point, associated with van Hove singularities, where the density of states is sharply enhanced.

\medskip
\par

\begin{figure}[h]
\centering
{\includegraphics[width=0.9\linewidth]{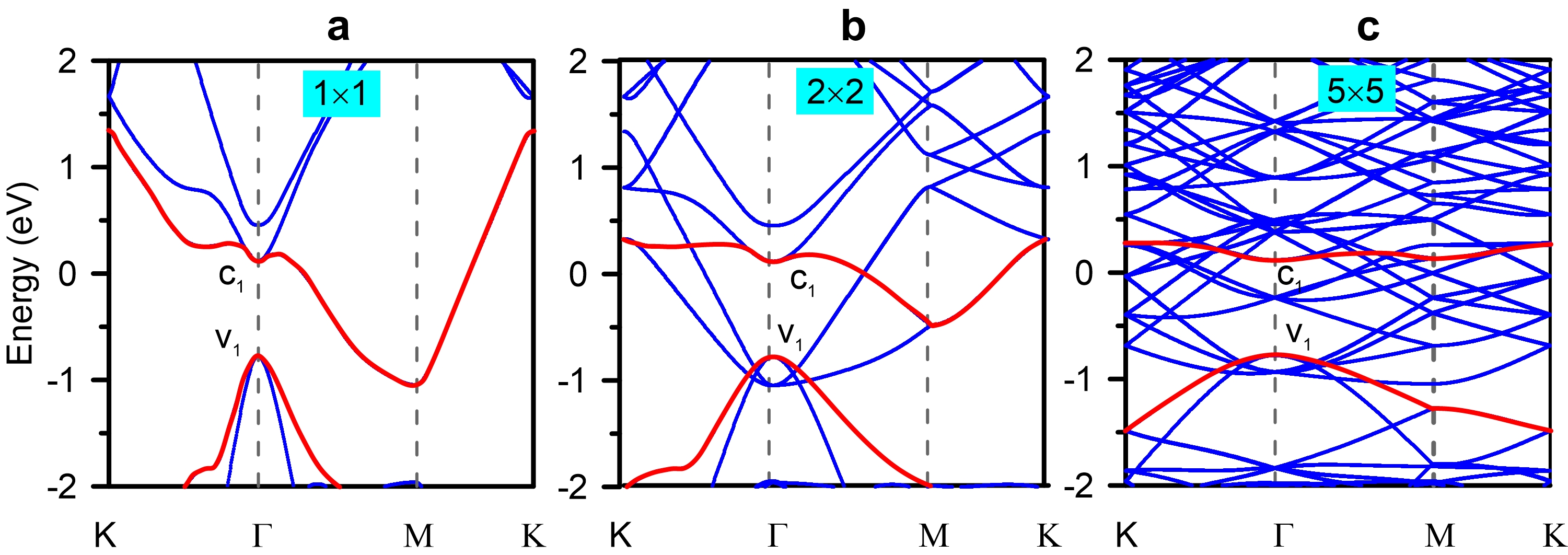}}
\caption{ {\bf a,} The band structure of TaS$_2$ at zero field. Under a weak honeycomb-lattice-shaped electric potential of 1 meV, the band structures are shown for {\bf b,} $2\times 2$ supercell and {\bf c,} $5\times 5$ supercell.}
\label{Fig2}
\end{figure}

An applied honeycomb-lattice-shaped potential modulation on TaS$_2$ is illustrated in Fig. 3. The potential is distributed in such a way that the maxima form a triangular lattice (blue color), while the minima create a honeycomb lattice (red color). This arrangement plays a crucial role in understanding the electronic structure and properties of the material. Previous studies on bilayer graphene have shown that honeycomb lattices tend to favor topological bands, whereas triangular lattices are more likely to host trivial bands \cite{bigraphene}. This demonstrates that lattice geometry is critical in determining the electronic characteristics of the material. Here, our numerical calculations for a weakly modulated potential (50 meV) on TaS$_2$ indicate that the conduction band electrons are localized near the potential minima, while the valence band holes are confined near the potential maxima.

\medskip
\par

\begin{figure}[h]
\centering
{\includegraphics[width=0.7\linewidth]{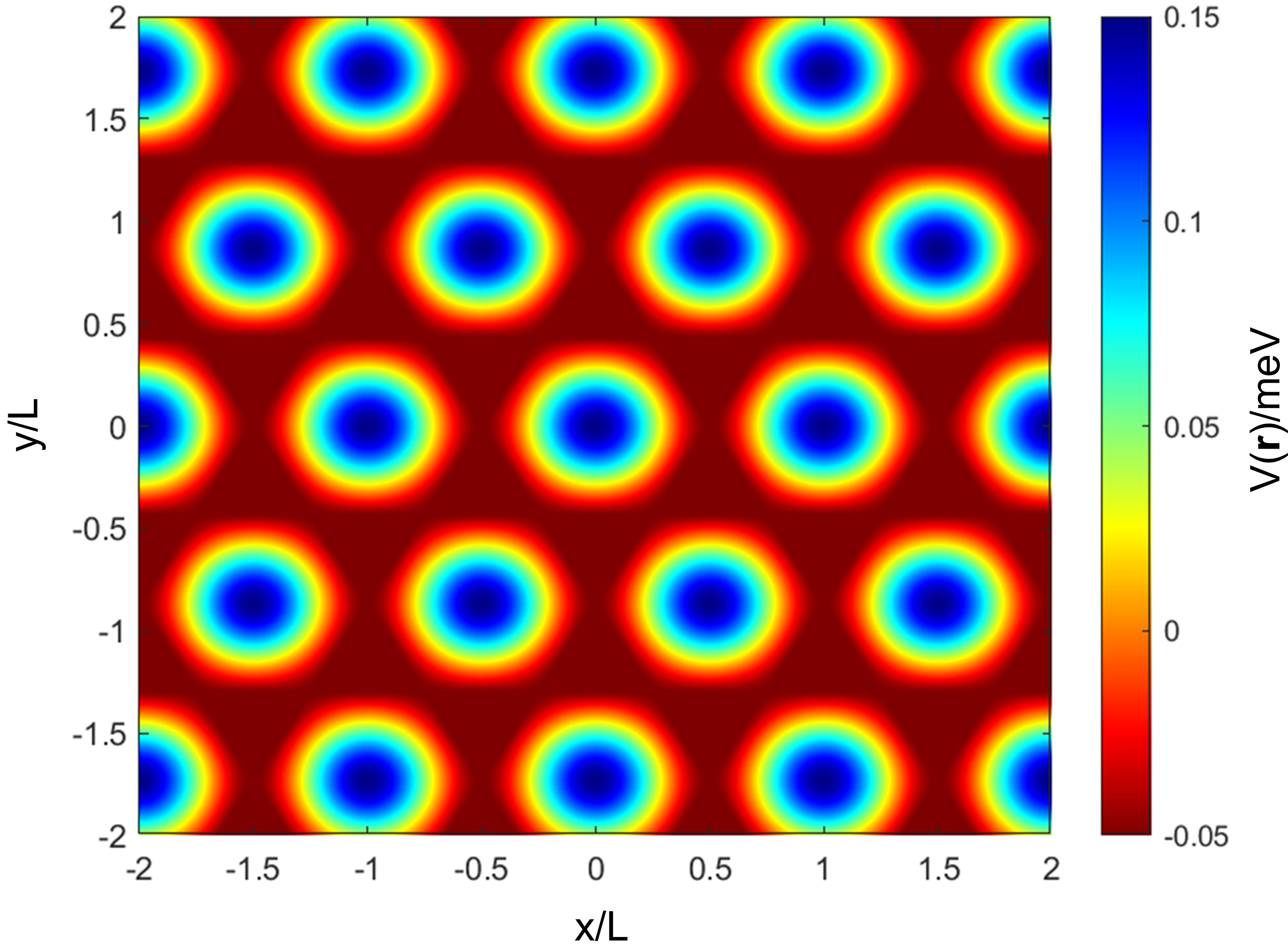}}
\caption{Superlattice potential with W = 50 meV.The potential maxima form a triangular lattice (blue color) while the potential minima form a honeycomb lattice (red color).}
\label{Fig3}
\end{figure}

When the lattice period is modified, band folding takes place in momentum space \cite{Sym2017Topp, bandfolding}. This process is accompanied by an expansion of the real-space unit cell and a reduction in the size of the first Brillouin zone (BZ). Specifically, if the primitive unit cell is doubled in size along the $x$ direction, the BZ will fold along the corresponding $k_x$ direction, causing the area of the BZ to be halved. In general, the more the unit cell is expanded, the greater the number of bands that are folded into the first BZ. This band folding results in the redistribution of the electronic states, which can significantly influence the material's electronic properties.

\medskip
\par

For TaS$_2$ under a superlattice potential modulation, band folding is induced and depends on the extent of the unit cell expansion. This phenomenon leads to the closing of the indirect band gap, an increase in the number of bands, and a modification of the band dispersion, as shown in Figs. 2(a) through 2(c). Notably, as the potential period increases, the pair of valence band $v_1$ and conduction band $c_1$ gradually flattens. In particular, for a $5 \times 5$ superlattice potential, we observe the emergence of a pair of flat bands, referring to Fig. 2(c).

\medskip
\par

\begin{figure}[h]
\centering
{\includegraphics[width=0.9\linewidth]{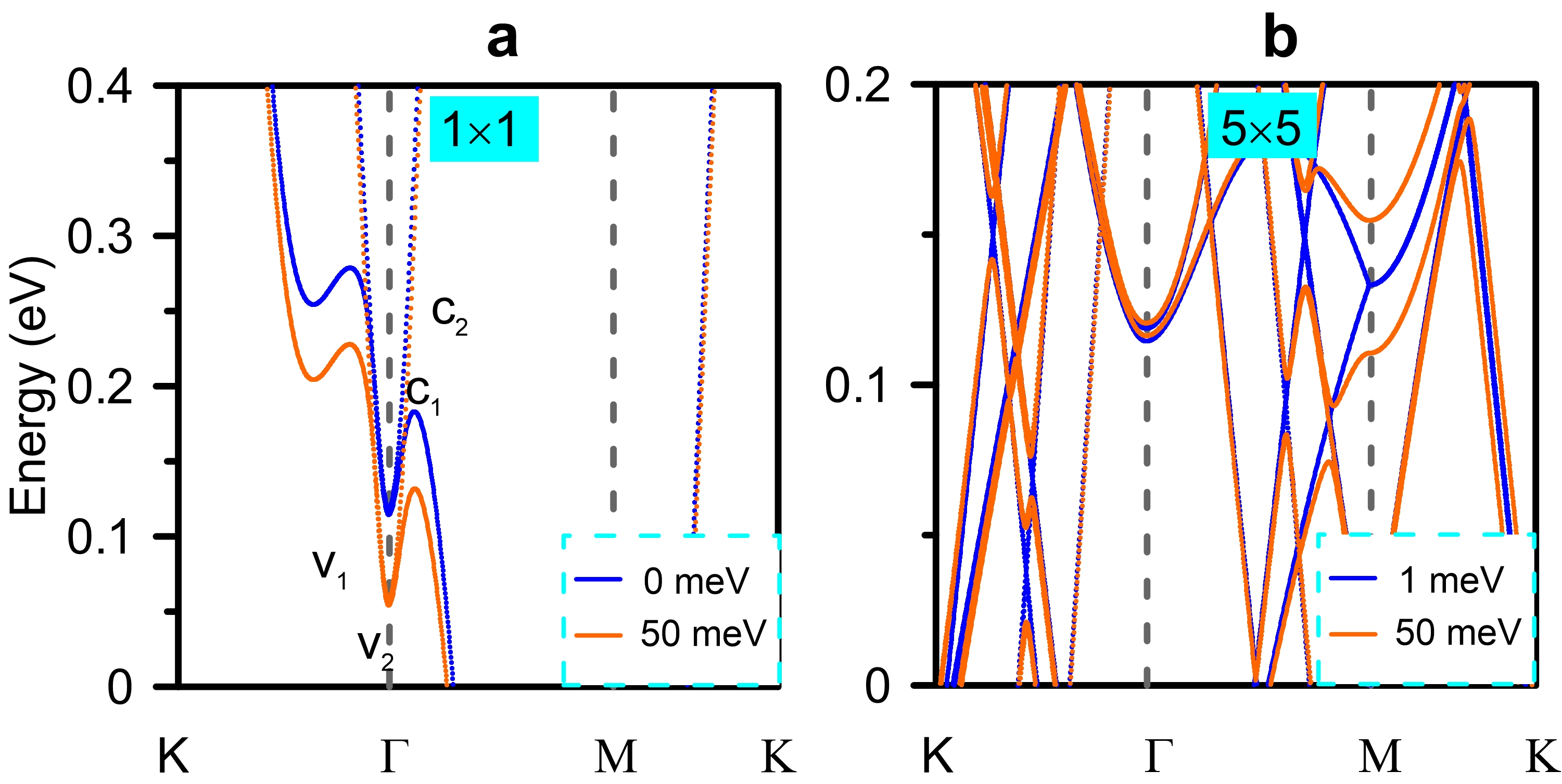}}
\caption{{\bf a,} The band structure of TaS$_2$ at zero field (blue curves) and at a $1\times 1$ superlattice potential of W = 50 meV (orange curves). {\bf b,} The band structures for $5\times 5$ supercell system under a superlattice potential of W = 1 meV (blue curves) and W = 50 meV (orange curves) are shown.}
\label{Fig4}
\end{figure}

The effects due to a modulated electric potential on the band structure and its dependence on the period  $n$ are illustrated in Figs. 4(a) and 4(b). When the potential period matches that of a single hexagon, there is no expansion of the unit cell, meaning no superlattice is formed. In this case, the potential causes only a slight shift of the energy bands, as shown in Fig. 4(a). However, when the unit cell expansion due to the modulated potential becomes significant, there is a noticeable interaction between neighboring bands, leading to more significant changes in the band structure. The band coupling results in the anticrossing of the bands involved as illustrated in Fig. 4(b). Theoretically, the anticrossing of energy bands occurs whenever two bands, initially close in energy, come near each other but avoid crossing due to the Pauli exclusion principle. This principle states that no two fermions, such as electrons, can occupy the same quantum state simultaneously. When two energy bands approach each other, the wavefunctions of the electrons in these bands begin to overlap, leading to a coupling between them. As a result, instead of crossing at the point where their energies would be identical, the bands split and bend around each other to open a band gap.

\medskip
\par

\section{Concluding Remarks}

In conclusion, we have developed a tight-binding model to investigate the electronic properties of TaS$_2$ under superlattice potential modulation. This model is derived from the Wannier tight-binding Hamiltonian, which are obtained using high-throughput density functional theory. Our computational approach allows for direct calculation of the band structures from the Hamiltonian, without relying on additional assumptions. We have observed the emergence of a pair of flat bands, significant interactions between energy bands, and a noticeable changes in band dispersion at low modulated electric potentials. This work provides valuable insights into the behavior of two-dimensional condensed matter materials under superlattice potential modulation and serves as a useful reference for future research in this area.

\medskip
\par

\section*{Acknowledgement(s)}
T.N.D.  would like to acknowledge the support from the University of Tampa. G.G. would like to acknowledge the support from the Air Force Research Laboratory (AFRL) through Grant No. FA9453-21-1-0046.

\medskip
\par


\begin{references}

\bibitem{QAHE} M. Serlin, C.L. Tschirhart, H. Polshyn, Y. Zhang, J. Zhu, K. Watanabe, T. Taniguchi, L. Balents, A.F. Young, Intrinsic quantized anomalous Hall effect in a moire heterostructure, Science 367 (2020) 900-903. https://doi.org/10.1126/science.aay5533.
    
\bibitem{QAHE1} A. Zhao, S.-Q. Shen, Quantum anomalous Hall effect in a flat band ferromagnet, Phys. Rev. B 85 (2012) 085209. https://doi.org/10.1103/PhysRevB.85.085209.

\bibitem{superconductor} X. Lu, P. Stepanov, W. Yang, M. Xie, M.A. Aamir, I. Das, C. Urgell, K. Watanabe, T. Taniguchi, G. Zhang, A. Bachtold, A.H. MacDonald, D.K. Efetov, Superconductors, orbital magnets and correlated states in magic-angle bilayer graphene, Nature 574 (2019) 653-657. https://doi.org/10.1038/s41586-019-1695-0.
    
\bibitem{superconductor1} H. Tian, X. Gao, Y. Zhang, S. Che, T. Xu, P. Cheung, K. Watanabe, T. Taniguchi, M. Randeria, F. Zhang, C.N. Lau, M.W. Bockrath, Evidence for Dirac flat band superconductivity enabled by quantum geometry, Nature 614 (2023) 440-444. https://doi.org/10.1038/s41586-022-05576-2.

\bibitem{charge} Y. Xie, B. Lian, B. Jack, X. Liu, C.-L. Chiu, K. Watanabe, T. Taniguchi, B.A. Bernevig, A. Yazdani, Spectroscopic signatures of many-body correlations in magic-angle twisted bilayer graphene, Nature 572 (2019) 101-105. https://doi.org/10.1038/s41586-019-1422-x.
    
\bibitem{charge1} H. Nakai, C. Hotta, Perfect flat band with chirality and charge ordering out of strong spin-orbit interaction, Nat Commun 13 (2022) 579. https://doi.org/10.1038/s41467-022-28132-y.

\bibitem{bigraphene} Y. Zeng, T.M.R. Wolf, C. Huang, N. Wei, S.A.A. Ghorashi, A.H. MacDonald, J. Cano, Gate-tunable topological phases in superlattice modulated bilayer graphene, Phys. Rev. B 109 (2024) 195406. https://doi.org/10.1103/PhysRevB.109.195406.
    
\bibitem{bigraphene1} S.A.A. Ghorashi, A. Dunbrack, A. Abouelkomsan, J. Sun, X. Du, J. Cano, Topological and Stacked Flat Bands in Bilayer Graphene with a Superlattice Potential, Phys. Rev. Lett. 130 (2023) 196201. https://doi.org/10.1103/PhysRevLett.130.196201.

\bibitem{gate1} D. Barcons Ruiz, H. Herzig Sheinfux, R. Hoffmann, I. Torre, H. Agarwal, R.K. Kumar, L. Vistoli, T. Taniguchi, K. Watanabe, A. Bachtold, F.H.L. Koppens, Engineering high quality graphene superlattices via ion milled ultra-thin etching masks, Nat Commun 13 (2022) 6926. https://doi.org/10.1038/s41467-022-34734-3.

\bibitem{gate2} D.Q. Wang, Z. Krix, O.P. Sushkov, I. Farrer, D.A. Ritchie, A.R. Hamilton, O. Klochan, Formation of Artificial Fermi Surfaces with a Triangular Superlattice on a Conventional Two-Dimensional Electron Gas, Nano Lett. 23 (2023) 1705-1710. https://doi.org/10.1021/acs.nanolett.2c04358.

\bibitem{exploi1} K. Yasuda, X. Wang, K. Watanabe, T. Taniguchi, P. Jarillo-Herrero, Stacking-engineered ferroelectricity in bilayer boron nitride, Science 372 (2021) 1458-1462. https://doi.org/10.1126/science.abd3230.

\bibitem{exploi2} D.S. Kim, R.C. Dominguez, R. Mayorga-Luna, D. Ye, J. Embley, T. Tan, Y. Ni, Z. Liu, M. Ford, F.Y. Gao, S. Arash, K. Watanabe, T. Taniguchi, S. Kim, C.-K. Shih, K. Lai, W. Yao, L. Yang, X. Li, Y. Miyahara, Electrostatic moire potential from twisted hexagonal boron nitride layers, Nat. Mater. 23 (2024) 65-70. https://doi.org/10.1038/s41563-023-01637-7.

\bibitem{application1} J. Bao, L. Yang, D. Wang, Influence of torsional deformation on the electronic structure and optical properties of 1T-TaS$_2$ monolayer, Journal of Molecular Structure 1258 (2022) 132667. https://doi.org/10.1016/j.molstruc.2022.132667.

\bibitem{application2} N.F. Hinsche, K.S. Thygesen, Electron-phonon interaction and transport properties of metallic bulk and monolayer transition metal dichalcogenide TaS2, 2D Mater. 5 (2017) 015009. https://doi.org/10.1088/2053-1583/aa8e6c.

\bibitem{cvp} X. Wang, H. Liu, J. Wu, J. Lin, W. He, H. Wang, X. Shi, K. Suenaga, L. Xie, Chemical Growth of 1T-TaS$_2$ Monolayer and Thin Films: Robust Charge Density Wave Transitions and High Bolometric Responsivity, Advanced Materials 30 (2018) 1800074. https://doi.org/10.1002/adma.201800074.

\bibitem{epitaxy} H. Lin, W. Huang, K. Zhao, S. Qiao, Z. Liu, J. Wu, X. Chen, S.-H. Ji, Scanning tunneling spectroscopic study of monolayer 1T-TaS2 and 1T-TaSe2, Nano Res. 13 (2020) 133-137. https://doi.org/10.1007/s12274-019-2584-4.

\bibitem{exfo} J. Peng, J. Wu, X. Li, Y. Zhou, Z. Yu, Y. Guo, J. Wu, Y. Lin, Z. Li, X. Wu, C. Wu, Y. Xie, Very Large-Sized Transition Metal Dichalcogenides Monolayers from Fast Exfoliation by Manual Shaking, J. Am. Chem. Soc. 139 (2017) 9019-9025. https://doi.org/10.1021/jacs.7b04332.

\bibitem{superconduct} C.-S. Lian, C. Heil, X. Liu, C. Si, F. Giustino, W. Duan, Coexistence of Superconductivity with Enhanced Charge Density Wave Order in the Two-Dimensional Limit of TaSe2, J. Phys. Chem. Lett. 10 (2019) 4076-4081. https://doi.org/10.1021/acs.jpclett.9b01480.

\bibitem{chargedensity} P.C. Borner, M.K. Kinyanjui, T. Bjorkman, T. Lehnert, A.V. Krasheninnikov, U. Kaiser, Observation of charge density waves in free-standing 1T-TaSe$_2$ monolayers by transmission electron microscopy, Applied Physics Letters 113 (2018) 173103. https://doi.org/10.1063/1.5052722.

\bibitem{ferro} J. He, S. Li, L. Zhou, T. Frauenheim, Ultrafast Light-Induced Ferromagnetic State in Transition Metal Dichalcogenides Monolayers, J Phys Chem Lett 13 (2022) 2765-2771. https://doi.org/10.1021/acs.jpclett.2c00443.

\bibitem{phase} H. Chen, J. Zhang, D. Kan, J. He, M. Song, J. Pang, S. Wei, K. Chen, The Recent Progress of Two-Dimensional Transition Metal Dichalcogenides and Their Phase Transition, Crystals 12 (2022) 1381. https://doi.org/10.3390/cryst12101381.

\bibitem{tas2flat} A. Dalal, J. Ruhman, J.W.F. Venderbos, Flat band physics in the charge-density wave state of 1T-TaS$_2$, (2024). https://doi.org/10.48550/arXiv.2406.18645.
    
\bibitem{tas2flat1} C. Wen, J. Gao, Y. Xie, Q. Zhang, P. Kong, J. Wang, Y. Jiang, X. Luo, J. Li, W. Lu, Y.-P. Sun, S. Yan, Roles of the Narrow Electronic Band near the Fermi Level in 1T-TaS$_2$-Related Layered Materials, Phys. Rev. Lett. 126 (2021) 256402. https://doi.org/10.1103/PhysRevLett.126.256402.
    
\bibitem{tas2flat2} L. Cheng, X. Long, X. Chen, X. Zou, Z. Liu, Understanding the flat band in 1T-TaS$_2$ using a rotated basis, Phys. Rev. B 104 (2021) L241114. https://doi.org/10.1103/PhysRevB.104.L241114.

\bibitem{jarvis} K.F. Garrity, K. Choudhary, Database of Wannier tight-binding Hamiltonians using high-throughput density functional theory, Sci Data 8 (2021) 106. https://doi.org/10.1038/s41597-021-00885-z.

\bibitem{tas2model} C. Tresca, M. Calandra, Charge density wave and spin 1/2 insulating state in single layer 1T-NbS2, 2D Mater. 6 (2019) 035041. https://doi.org/10.1088/2053-1583/ab23c0.

\bibitem{folding} A. Mrenca-Kolasinska, S.-C. Chen, M.-H. Liu, Probing miniband structure and Hofstadter butterfly in gated graphene superlattices via magnetotransport, Npj 2D Mater Appl 7 (2023) 1-9. https://doi.org/10.1038/s41699-023-00426-9.

\bibitem{Hsl} T. Li, J. Ingham, H.D. Scammell, Artificial graphene: Unconventional superconductivity in a honeycomb superlattice, Phys. Rev. Res. 2 (2020) 043155. https://doi.org/10.1103/PhysRevResearch.2.043155.

\bibitem{Sym2017Topp} A. Topp, R. Queiroz, A. Gruneis, L. Muchler, A.W. Rost, A. Varykhalov, D. Marchenko, M. Krivenkov, F. Rodolakis, J.L. McChesney, B.V. Lotsch, L.M. Schoop, C.R. Ast, Surface Floating 2D Bands in Layered Nonsymmorphic Semimetals: ZrSiS and Related Compounds, Phys. Rev. X 7 (2017) 041073. https://doi.org/10.1103/PhysRevX.7.041073.

\bibitem{bandfolding} B.I. Min, Y.-R. Jang, Band folding and Fermi surface in antiferromagnetic ${\mathrm{NdB}}_{6}$, Phys. Rev. B 44 (1991) 13270-13276. https://doi.org/10.1103/PhysRevB.44.13270.


\end{references}
\end{document}